\documentclass{aa}  

\usepackage{graphicx}
\usepackage{txfonts}
\usepackage{xcolor}
\usepackage{soul}
\usepackage{hyperref}
\usepackage{mathrsfs}
\usepackage{amsmath}

\begin{document} 

   \title{Relative evolution of eclipsing binaries:\\A tool to measure globular cluster ages and He abundances}

   \author{N. Cristi-Cambiaso\inst{1, 2}
          \and
          M. Catelan\inst{1, 2, 3}
          \and
          A. A. R. Valcarce\inst{2, 4}
          \and
          A. Papageorgiou\inst{5}
          }

   \institute{
            Instituto de Astrofísica, Pontificia Universidad Católica de Chile, Av. Vicuña Mackenna 4860, 7820436 Macul, Santiago, Chile\\ \email{[nicolascristi;mcatelan]@uc.cl}
            \and
            Millennium Institute of Astrophysics, Nuncio Monseñor Sotero Sanz 100, Of. 104, Providencia, Santiago, Chile
            \and
            Centro de Astro-Ingeniería, Pontificia Universidad Católica de Chile, Av. Vicuña Mackenna 4860, 7820436 Macul, Santiago, Chile
            \and
            Departamento de Física, FACI, Universidad de Tarapacá, Casilla 7D, Arica, Chile
            \and
            Department of Physics, University of Patras, 26500, Patra, Greece
             }

   \date{Received April 17, 2024; accepted September 4, 2024}

 
  \abstract
   {Globular clusters (GCs) are among the oldest objects in the Universe for which an age can be directly measured, thus playing an important cosmological role. This age, on the other hand, depends sensitively on the He abundance, which cannot be reliably measured from spectroscopy in GC stars. Detached eclipsing binaries (DEBs) near the turnoff (TO) point may play an important role in this regard.}
   {The aim of this study is to explore the possibility that, by working with differential measurements of stars that comprise a TO binary system, and assuming both stars have the same age and He abundance, one can achieve tighter, more robust, and less model-dependent constraints on the latter two quantities than otherwise possible by working with the absolute parameters of the stars.}
   {We compare both absolute and differential parameters of the stars in V69, a TO DEB pair in the GC 47~Tuc, with two different sets of stellar evolutionary tracks, making use of a Monte Carlo technique to estimate its He abundance and age, along with their uncertainties.}
   {We find that the relative approach can produce age and He abundance estimates that are in good agreement with those from the literature. We show that our estimates are also less model-dependent, less sensitive to [Fe/H], and more robust to inherent model systematics than those obtained with an absolute approach. On the other hand, the relative analysis finds larger statistical uncertainties than does its absolute counterpart, at least in the case of V69, where both stars have very similar properties. 
   For binary pairs in which one of the components is less evolved than the other, the statistical uncertainty can be reduced.
   }
   {Our study suggests that the method proposed in this work may be useful to robustly constrain the He abundance and ages of GCs.}

   \keywords{stars: abundances - stars: evolution - globular clusters: individual (47\,Tuc) - stars: individual (V69-47\,Tuc) - (stars:) binaries: eclipsing}

   \maketitle
%

\section{Introduction}
\label{sec:introduction}

Detached eclipsing binary (DEB) systems are exceptional subjects for astronomical studies due to their orbits, which enable a direct determination of their masses, radii, and other stellar parameters \citep[see][for a detailed explanation]{Paczynski1997}. This direct determination is achieved through the analysis of their photometric light curves and radial velocities, reaching errors in the stellar parameters as small as $1$\% in some cases \citep{Thompson2020}. Furthermore, when at least one star within a DEB system is located at (or very close to) the main sequence (MS) turnoff (TO) point, it becomes a valuable tool for accurately estimating ages, since the TO is a key age indicator in the realm of stellar evolution (see, e.g., the reviews by \citeauthor{Soderblom2010} \citeyear{Soderblom2010} and \citeauthor{Catelan2018} \citeyear{Catelan2018}, and references therein). The use of these DEBs for age dating of a stellar cluster was initially proposed by \citet{Paczynski1997}, and has since become a common practice in numerous studies \citep[e.g.,][]{Grundahl2008, Brogaard2011}. Several DEBs in open and globular clusters (GCs), in which at least one component is located near the TO point, are now known \citep[see][and references therein]{Rozyczka2022}.

One especially interesting system is the double-lined DEB V69-47 Tuc (hereafter, V69), located within the GC 47\,Tucanae (47\,Tuc, NGC\,104). This binary system consists of two (very similar) low-mass stars that are both near the TO point, making V69 an excellent candidate for determining 47\,Tuc's age. The significance of V69 in determining the age of 47\,Tuc has not gone unnoticed, as it has been extensively utilized with this purpose in multiple studies \citep[e.g.,][]{Dotter2009, Brogaard2017, Thompson, Thompson2020}. 
When attempting to match the empirically derived physical parameters of the V69 components, however, a problem arises, as the consideration of different helium abundances ($Y$, the fraction of He by mass) in the stellar models introduces a wide range of possible ages for the system, spanning several billion years. In other words, a strong degeneracy arises between the assumed $Y$ of the binary and its estimated age. This poses a challenge, since it is not feasible to directly determine $Y$ for moderately cool stars, such as the components of V69, given their inability to excite He atoms in their photospheres. Since V69 is a cluster member, one could in principle use measurements for other cluster stars for this purpose; however, the only stars for which this can be done directly are blue horizontal-branch stars, whose photospheric abundances are severely affected by diffusion, and whose measured He abundances are thus not representative of the cluster's initial $Y$ \citep[e.g.,][and references therein]{Michaud2015}. While the chromospheric He\,{\sc i} line at 10,830\,\AA\ may in principle provide a direct measurement \citep{Dupree2013}, the technique is still far from being able to yield representative $Y$ values from equivalent width measurements.

A further complication that arises when comparing evolutionary tracks and isochrones with the empirical data is that stellar models are inherently affected by systematic uncertainties, as often are empirical measurements as well. These uncertainties encompass various factors, such as color-temperature transformations, boundary conditions, treatment of turbulent convection, etc. \citep[e.g.,][and references therein]{Catelan2013,Pisa2013a,Cassisi2016,Buldgen2019}. Such systematic uncertainties constitute the main reason why the so-called ``horizontal methods'' of age-dating GCs \citep{Sarajedini1990, Vandenberg1990} are used to derive ages in a differential sense only \citep[see also][]{Chaboyer1996,Sarajedini1997,VandenBerg2013,Catelan2018}. In fact, these problems affect traditional isochrone-fitting methods as well, as pointed out early on by \citet{Iben1984}, among many others. For this reason, whenever the temperatures, colors, and radii of stars are involved, differential analyses are expected to give more reliable results than absolute ones.

Our main hypothesis is that, at least in principle, dealing with the binary components {\em differentially} can help minimize the impact of systematic errors affecting the empirical measurements and theoretical models alike. Indeed, several studies have demonstrated that the difference in temperature between eclipsing binary members can be derived with higher precision than is possible from the absolute temperatures of its individual components \citep[e.g.,][]{Southworth2007, Torres2014, Taormina2024}. In addition, this
rationale aligns with the approach employed in the study of solar twins and the differential analyses commonly conducted in these cases \citep[][]{Melendez2014, Spina2018, Martos2023}, which can yield remarkably precise stellar parameters: in the case of effective temperature ($T_{\text{eff}}$), precision at the level of $\sim 3$\,K has been reported.

In this paper, we explore the possibility that, by working with differential measurements of the properties of the stars that comprise a binary pair, and assuming both stars have the same age and $Y$, one may be able to achieve more robust constraints on the latter two quantities than would have been possible by working with both stars' absolute parameters. In this first experiment, we employ V69 in 47~Tuc. 
In Sect.~\ref{sec:StellarParameters}, we detail the stellar parameters and chemical composition that we have adopted for the analysis of V69. Sect.~\ref{sec:models} describes the theoretical models used in our study, while Sect.~\ref{sec:methods} outlines the computation of relative parameters for both the observed stars and the models, and the fitting procedure employed to derive an age and $Y$ for V69. Section~\ref{sec:results} presents the results obtained with our method for V69, followed by a summary of this work, in Sect.~\ref{sec:conclusions}.

\section{Adopted stellar parameters and chemical composition of V69}
\label{sec:StellarParameters}

We adopt the masses ($M$), bolometric luminosities ($L_{\text{bol}}$), and $T_{\text{eff}}$ of V69's components from \citet{Thompson2020}. Their respective radii ($R$) and surface gravities ($g$) come from \citet{Thompson}, as revised estimates are not provided in \citet{Thompson2020}. These values, based on a combination of photometry and spectroscopy, are given in Table~\ref{tab:V69}. Additionally, Table~\ref{tab:V69} shows the relative values of each parameter with their respective errors; these are explained in Sect.~\ref{sec:diff_params}.
In this work, the iron-over-hydrogen ratio with respect to the Sun, [Fe/H], and $\alpha$-element enhancement of the system, [$\alpha$/Fe], are assumed to be [Fe/H] = $-0.71 \pm 0.05$ and [$\alpha$/Fe] = $+0.4$, respectively. This chemical composition was adopted based on the studies conducted by \citet{Thompson, Thompson2020} and \citet{Brogaard2017}. The adopted [Fe/H] value agrees, to within $0.01$~dex, with the value obtained by \citet{Roediger2014} from a systematic compilation of measurements in the literature, and also with the [Fe/H] value recently recommended by \citet{VandenBerg2024}.

\begin{table*}
    \centering
    \caption{\footnotesize Absolute and relative stellar parameters of the V69 system.}
    \begin{tabular}{l l l l}
        \hline
        Parameter & Secondary & Primary & Relative \\
        \hline
        $M (M_{\odot})$ & $0.8584 \pm 0.0042$ & $0.8750 \pm 0.0043$ & $-0.0166 \pm 0.0060$ \\ 
        $R (R_{\odot})$ & $1.1616 \pm 0.0062$ & $1.3148 \pm 0.0051$ & $-0.153 \pm 0.008$ \\
        $T_{\text{eff}}$ (K) & $5988 \pm 46$ & $5959 \pm 45$ & $29 \pm 64$ \\
        $L_{\text{bol}} (L_{\odot})$ & $1.56 \pm 0.05$ & $1.96 \pm 0.06$ & $-0.40 \pm 0.08$ \\
        $\log g$ (dex) & $4.242 \pm 0.003$ & $4.143 \pm 0.003$ & $0.0990 \pm 0.0042$ \\
        \hline
    \end{tabular}
    \label{tab:V69}
    \tablefoot{The absolute parameters were obtained by \citet{Thompson} and \citet{Thompson2020}. The nominal errors in the relative parameters were obtained by propagating the errors in the absolute ones.}
\end{table*}

\section{Theoretical stellar models}
\label{sec:models}

We compare the physical parameters of V69 with stellar models obtained from two stellar evolution codes (SECs): Princeton-Goddard-PUC \citep[PGPUC,][]{PGPUC2012, PGPUC2013} and Victoria-Regina \citep[][referred to as VR hereafter]{VandenBerg2014}. These models were chosen because they offer a wide range of He abundances and metallicities, as can be seen in Table~\ref{tab:Tracks}. We obtain evolutionary tracks with different mass values than those available in the original grids by interpolation. Further details regarding the latter are provided in Appendix~\ref{app:interpolation}.

The PGPUC tracks provided by \citet{PGPUC2012, PGPUC2013} start at the zero-age main sequence (ZAMS) and extend to the He flash, at the tip of the red giant branch (RGB). 
For our analysis, we compute tracks from the PGPUC Online webpage\footnote{\url{http://www2.astro.puc.cl/pgpuc/index.php}} with the masses, He abundances, and metallicities specified in Table~\ref{tab:Tracks}. 
As mentioned in Sect.~\ref{sec:StellarParameters}, the [$\alpha$/Fe] of V69 is adopted as $+0.4$. However, we use PGPUC models with [$\alpha$/Fe] = $+0.3$, since \citet{PGPUC2012, PGPUC2013} do not offer models with a higher [$\alpha$/Fe]. 
The impact of this $0.1$~dex difference upon our results will be addressed later in the paper.

\citet{VandenBerg2014} provide grids of evolutionary tracks that start at the ZAMS and extend up to the tip of the RGB, in addition to computer programs that enable interpolation of new models within these grids \citep[see][for more details on the interpolating programs]{VandenBerg2014}. However, these programs do not support interpolation of tracks with new masses (only new chemical compositions), and the mass resolution of the available grid is insufficient for our purposes. Therefore, we employ these programs to interpolate new isochrone grids (their ages and chemical compositions are shown in Table~\ref{tab:Tracks}), which we subsequently use to interpolate tracks (see Appendix~\ref{app:interpolation}) with the same masses as the tracks in the PGPUC grid.

Both the PGPUC and VR databases directly provide the values for $ T_{\text{eff}}$, $L_{\text{bol}}$, and $M$, among others, for each model. However, the corresponding $R$ values are not provided, and, in the VR case, neither are the $g$ values. Hence, the radii are calculated according to the Stefan-Boltzmann law,  
$R = [L_{\text{bol}}/(4 \pi \sigma T_{\text{eff}}^4)]^{1/2}$, where $\sigma$ is the Stefan-Bolztmann constant, while the $g$ values are calculated from Newton's law of universal gravitation, $g = G M / R^2$, where $G$ is the gravitational constant.

\begin{figure*}
    \includegraphics[width=2\columnwidth]{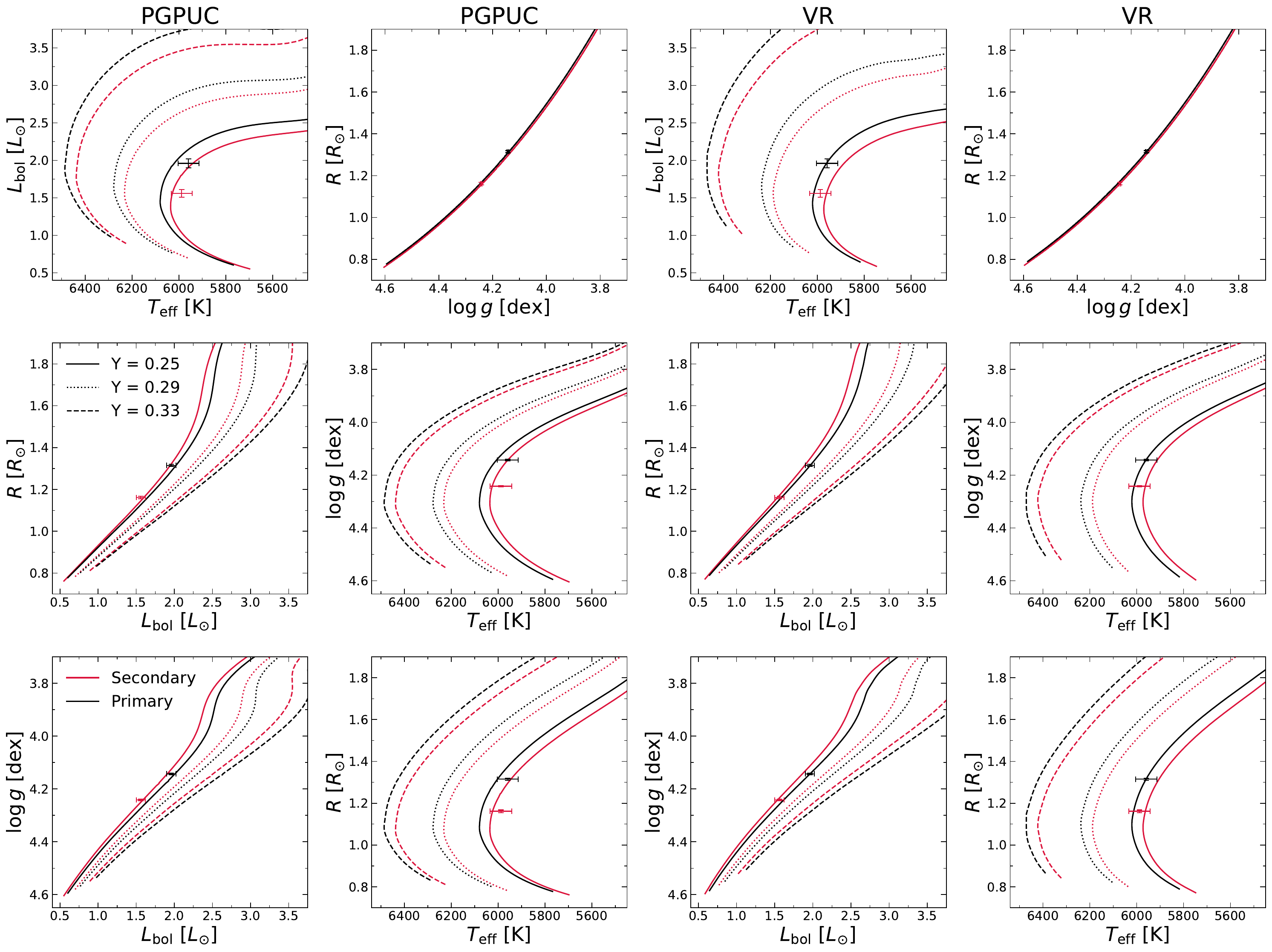}
    \caption{Evolutionary tracks from PGPUC (first and second columns) and VR (third and fourth columns) with the \textcolor{red}{nominal} masses of V69's components (\textcolor{red}{from Table~\ref{tab:V69}}; red tracks correspond to the secondary, while black tracks correspond to the primary), compared with the observations (Table~\ref{tab:V69}) of the binary. Solid, dotted, and dashed lines correspond to $Y = 0.25$, $Y = 0.29$, and $Y = 0.33$, respectively. All models were computed with [$\alpha$/Fe] = $0.3$. The VR tracks have $\text{[Fe/H]} = -0.7$, while the PGPUC tracks have $Z = 0.0057$, which at $Y = 0.25$ corresponds to $\text{[Fe/H]} = -0.7$. The VR tracks are interpolated following the procedure explained in Appendix~\ref{app:interpolation}.}
    \label{fig:absolute_tracks}
\end{figure*}

\begin{figure*}
    \includegraphics[width=2\columnwidth]{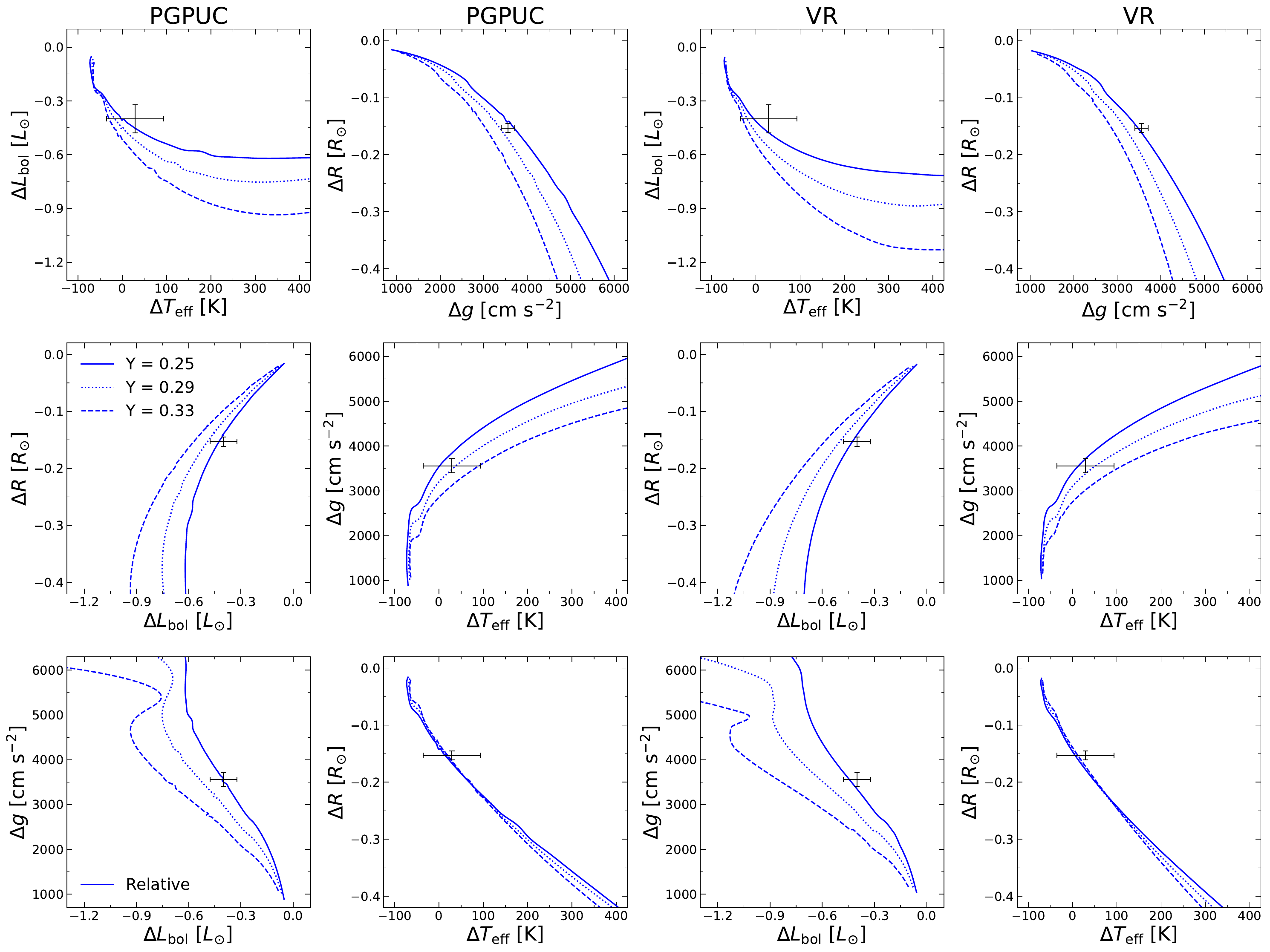}
    \caption{Same as Fig.~\ref{fig:absolute_tracks}, but displaying the relative parameters of the tracks and stars, instead of the absolute ones. Since the absolute mass values are fixed according to the nominal values given in Table~\ref{tab:V69}, tracks for different $Y$ values have different TO ages.}
    \label{fig:relative_tracks}
\end{figure*}

Figure~\ref{fig:absolute_tracks} overplots the absolute stellar parameters of the V69 pair (from Table~\ref{tab:V69}) on evolutionary tracks interpolated from the PGPUC and VR grids for different He abundances. These tracks are computed for the nominal mass values of both stars (also from Table~\ref{tab:V69}). 
Additionally, Fig.~\ref{fig:relative_tracks} shows a comparison between the relative parameters of the system and those predicted by these tracks (obtained as described in the next section). 
These plots show that, in the absolute and general cases alike, there is general agreement between both sets of tracks and the data, with values of $Y$ at the lower range of those explored apparently being favored.\footnote{Though not displayed in these figures, we have checked that Bag of Stellar Tracks and Isochrones (BaSTI) models \citep{Pietrinferni2021} for a similar chemical composition ($Y = 0.255$, [Fe/H] = $-0.7$, [$\alpha$/Fe] = $0.4$, $Z = 0.006$) match closely the PGPUC and VR models in the relevant region of parameter space.} This will be explored further in the following sections.

\begin{table*}
    \caption[Track parameters]{\footnotesize Description of the PGPUC and VR theoretical models utilised in this work.}
    \centering
    \begin{tabular}{l l l l} 
        \hline
        Parameter & PGPUC\tablefootmark{a} & VR\tablefootmark{b} \\
        \hline
        Mass ($M_{\odot}$) & $0.84$ to $0.90$ ($\delta = $$0.0005$)\tablefootmark{c} & ... \\
        Age (Gyr) & ... & $1$ to $5$ ($\delta = 1$), $5$ to $18$ ($\delta = 0.1$) \\
        $Y$ & $0.23$ to $0.37$ ($\delta = 0.005$) & $0.25$ to $0.33$ ($\delta = 0.005$) \\
        $Z$ & $0.0052$, $0.0060$, $0.0068$, $0.0076$ & ... \\
        {[}Fe/H{]} & $-0.74$ to $-0.57$ \tablefootmark{d} & $-0.75$, $-0.7$, $-0.65$, $-0.6$ \\
        {[}$\alpha$/Fe{]} & $+0.3$ & $+0.3$, $0.35$, $+0.4$ \\
        Solar mixture & (1) & (2) \\
        \hline
        \end{tabular}
        \label{tab:Tracks}
        \tablebib{(1) \citet{Grevesse1998}; (2) \citet{Asplund2009}.}
        \tablefoot{
        \tablefoottext{a}{Evolutionary tracks are used in this case.}
        \tablefoottext{b}{Isochrones are used in this case (see Sect.~\ref{sec:models}).}
        \tablefoottext{c}{$\delta$~= step size.}
        \tablefoottext{d}{Calculated based on the listed $Z$, $Y$, and [$\alpha$/Fe] values, using a tool provided on the PGPUC Online webpage, \url{http://www2.astro.puc.cl/pgpuc/FeHcalculator.php}.}
        }
\end{table*}

\section{Methods}
\label{sec:methods}

In this section, we describe the method proposed in this work to estimate the He abundance and age of the binary V69. The methods are illustrated by means of PGPUC calculations carried out for a single metallicity, $Z = 0.006$. A more detailed analysis that considers the range of possible metallicities of the system and incorporates VR models is presented in Sect.~\ref{sec:results}.

\subsection{Relative parameters}
\label{sec:diff_params}

To perform our differential analysis, we compute the relative physical parameters of the binary $\Delta \mathscr{X}$, where $\mathscr{X}$ may represent any of $M$, $T_{\rm{eff}}$, $L_{\rm{bol}}$, $g$, and $R$, as follows: 

\begin{equation}
    \Delta \mathscr{X}(t) = \mathscr{X}_\text{s}(t) - \mathscr{X}_\text{p}(t),
    \label{eq: diff}
\end{equation}

\noindent where $\mathscr{X}_{\rm s}(t)$ and $\mathscr{X}_{\rm p}(t)$ are the physical parameters of the secondary and primary, respectively, computed for a given age $t$, the latter assumed identical for both members of the binary system.

The nominal errors in these relative parameters are obtained by propagating those in the absolute parameters in quadrature. However, as mentioned in Sect.~\ref{sec:introduction}, since the components of V69 are very similar, they should be affected similarly by systematic uncertainties (due to, e.g., the use of different color indices, bolometric corrections, color-temperature transformations, etc.). For a strictly differential analysis, therefore, we posit that systematic errors affecting each of the two components individually do {\em not} propagate in quadrature as statistical errors do; rather, they should largely cancel each other out. 
We thus consider that the errors propagated in quadrature are upper limits. This is explored further in Sect.~\ref{sec:precision}.

In like vein, a differential evolutionary track was obtained for the binary system, by computing the difference, for each age $t$, between the physical parameters of the secondary and primary along their respective tracks. This was done by interpolating in time within the individual tracks, which are uniquely defined by the specified $(M_{\rm s}, Z_{\rm s}, Y_{\rm s})$, $(M_{\rm p}, Z_{\rm p}, Y_{\rm p})$ combinations, with (in general) $M_{\rm s} < M_{\rm p}$, metallicity $Z_{\rm s} = Z_{\rm p}$, and helium abundance $Y_{\rm s} = Y_{\rm p}$. Some such differential tracks are shown in Fig.~\ref{fig:relative_tracks}.

An alternative approach to the one used in this paper is to employ parameter {\em ratios}, as opposed to differential parameters. This is motivated by the fact that, when studying binary systems, such ratios are often more directly measured from the empirical data \citep[e.g.,][]{Popper1985, Maxted2016, Morales2022}. 
We explore this possibility further in Appendix~\ref{app:ratios}.

\subsection{Measuring track-to-star agreement}

To estimate the $Y$ value and age of the binary, we compare its (absolute and relative) parameters with a diverse set of stellar models in a four-dimensional space of $M$, $L_{\text{bol}}$, $g$, and $R$. Effective temperature is included implicitly, through the Stefan-Boltzmann relation that relates it to $L_{\text{bol}}$ and $R$. We do not include $T_{\rm eff}$ explicitly because, as can be seen in Fig.~\ref{fig:relative_tracks}, it has less diagnostic power than other physical parameter combinations, in the relative case.

To evaluate the quality of the fits for different ($Y$, age) combinations, we define a goodness-of-fit parameter as follows. 
We calculate Euclidean track-to-star distances (TSDs), 

\begin{equation}\label{eq:TSD}
\begin{aligned}
\text{TSD}(t) = \left\{\left[\frac{M_{\star} - M_{\rm ET}(t)}{\sigma_{M_{\star}}}\right]^2 + \left[\frac{L_{\rm{bol}, \star} - L_{\rm{bol, ET}}(t)}{\sigma_{L_{\rm{bol},\star}}}\right]^2 \right.\\
\left. + \left[\frac{R_{\star} - R_{\text{ET}}(t)}{\sigma_{R_\star}}\right]^2 + \left[\frac{g_{\star} - g_{\text{ET}}(t)}{\sigma_{g_\star}}\right]^2\right\}^{1/2}, 
\end{aligned}
\end{equation}

\noindent where $M_{\text{ET}}(t)$, $L_{\rm{bol,ET}}(t)$, $R_{\text{ET}}(t)$, and $g_{\text{ET}}(t)$ are the evolutionary track's mass, bolometric luminosity, radius, and surface gravity at time $t$, respectively; $M_{\star}$, $L_{{\rm bol},\star}$, $R_{\star}$, $g_{\star}$ the corresponding empirical values; and $\sigma_{M_{\star}}$, $\sigma_{L_{{\rm bol},\star}}$, $\sigma_{R_{\star}}$, and $\sigma_{g_{\star}}$  their respective uncertainties. Dividing each term by the error of the respective parameter gives higher weight to those parameters that are empirically better constrained, and renders TSD a quantity with no physical units~--- an approach similar to the one followed, for instance, by \citet{Pols1997} and \citet{Joyce2023}.

The TSDs were calculated for both the relative (RTSDs) and absolute (ATSDs) approaches. The RTSDs are simply the TSDs computed for a relative track (such as those shown in Fig.~\ref{fig:relative_tracks}) and the ``relative star'', i.e., the relative parameters of the binary system. On the other hand, the ATSDs are computed as 

\begin{equation}
    \text{ATSD}(t) = \sqrt{\text{TSD}_{\text{p}}^2(t) + \text{TSD}_{\text{s}}^2(t)} \, ,
\end{equation}

\noindent where $\text{TSD}_{\text{s}}(t)$ and $\text{TSD}_{\text{p}}(t)$ are the TSDs computed for the secondary and primary track, respectively.

\subsection{Helium abundance and age determination}
\label{sec:Y_age}

For each set of evolutionary models (i.e., PGPUC and VR), we find all the ($M_{\rm s}$, $M_{\rm p}$) combinations, from the available masses in the model grid (see Table~\ref{tab:Tracks}), that satisfy $M_{\rm s} < M_{\rm p}$, discarding those instances in which the masses are outside $3 \sigma$ of the respective star's mass. Then, for each of these mass combinations, we compute relative tracks for every available He abundance value (as explained in Sect.~\ref{sec:diff_params}), and calculate ATSD and RTSD values for all of the tracks. Given the strong dependence of the relative tracks on mass around the MS TO point, a high enough mass resolution is required in order to infer reliable $Y$ and age values using this method.

Finally, we search through all the mass combinations to find the evolutionary track(s) that are associated with the minimum ATSD and RTSD values, here denoted MATSD and MRTSD, respectively. These values will correspond to the point in the evolutionary track(s) that comes closest to the measured binary parameters (i.e. $M$, $L_{\text{bol}}$, $R$, and $g$), according to the adopted statistic. The estimated $Y$ and age of the binary can then be straightforwardly obtained from this closest point of the track(s).

\subsection{Accounting for uncertainties}
\label{sec:iterations}
The nominal $M$, $L_{\text{bol}}$, $R$, and $g$ values of the stars, as adopted in Sect.~\ref{sec:Y_age}, are not necessarily the {\em true} ones. As shown in Table~\ref{tab:V69}, V69's parameters also have an associated uncertainty, which must thus be taken into account. Here we employ a Monte Carlo (MC) technique to perform an analysis of the impact of the errors in these parameters. 
We run $2000$ MC instances, in which we resample the $M$, $L_{\text{bol}}$, $R$, and $g$ of both stars assuming they follow normal deviates. The latter follow the binary components' nominal values and associated errors, both from Table~\ref{tab:V69}, as their mean and standard deviation, respectively. The only constraint imposed is that the primary's mass must be greater than the secondary's. 
In each MC instance, we then follow the same procedure described earlier to estimate anew the $Y$ value and age of the binary system, adopting these resampled parameters. 
This sampling procedure assumes that the uncertainties in the relative parameters are not correlated. This notwithstanding, and as will be discussed below (Sect.~\ref{sec:testing}), our artificial star tests show that the method is able to recover the correct age and $Y$ values. In future applications, however, the possibility of correlated errors in at least some of the input parameters should be investigated and taken into account if present, in order for more realistic errors in the inferred parameters to be extracted from the data.

\begin{figure}
    \centering
    \includegraphics[width=1\columnwidth]{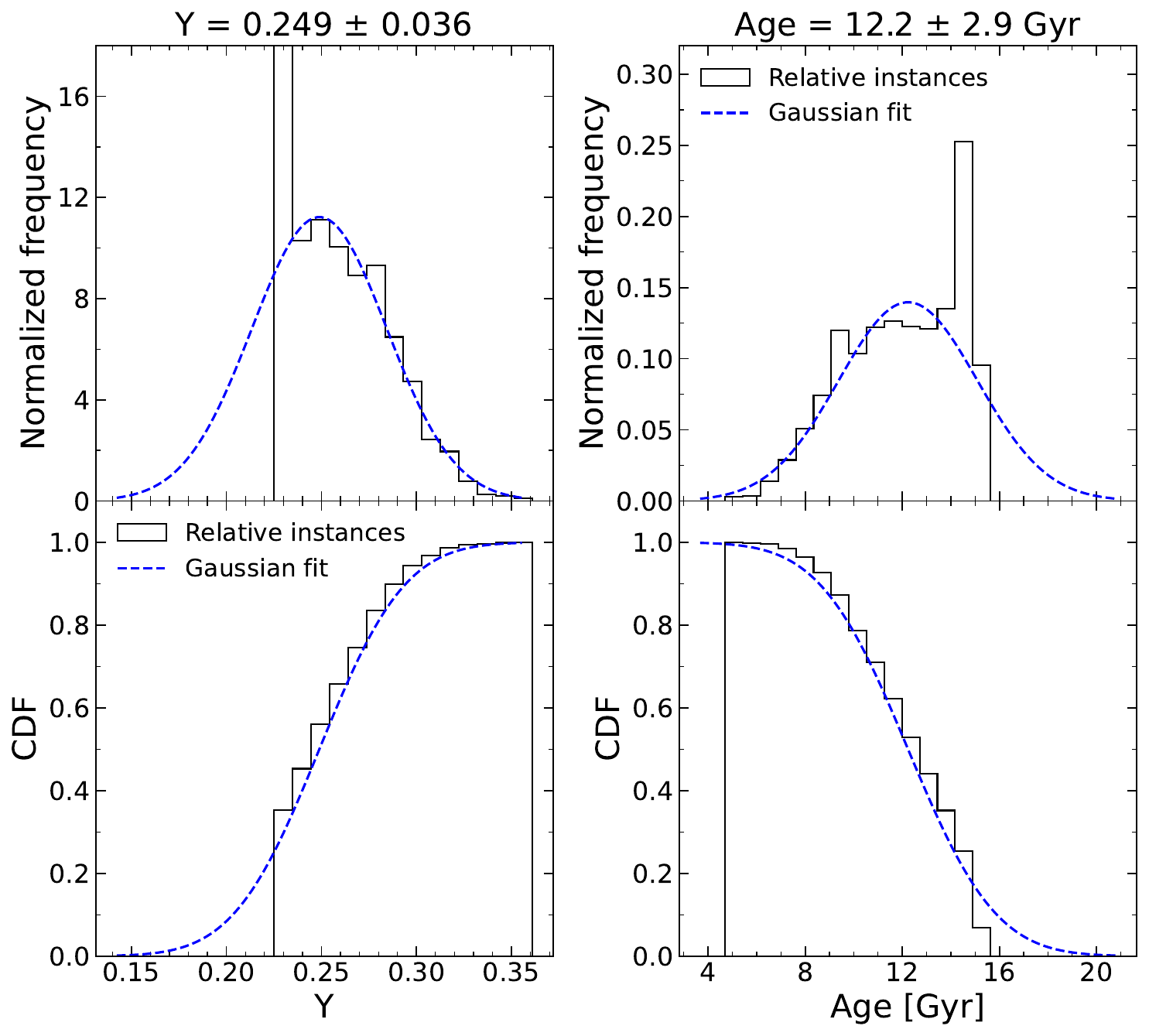}
    \caption{$Y$ (left) and age (right) distributions across $2000$ MC instances, using relative PGPUC tracks with $Z = 0.006$. The upper panels show the $Y$ and age distributions, while the lower panels show the corresponding CDFs. The best-fitting Gaussian CDFs are shown in the lower panels (as dashed blue lines), with their mean and standard deviation displayed in the title of the upper panels. Additionally, Gaussians crafted with these parameters are shown in the upper panels (as dashed blue lines). The lowest $Y$ bin in the upper left panel, which is not shown completely, has a normalized frequency of $37$.}
    \label{fig:iterations_rel}
\end{figure}

Figure~\ref{fig:iterations_rel} shows the $Y$ and age distributions obtained when using relative PGPUC models with $Z = 0.006$. As shown in the upper left panel, the $Y$ distribution can present an overpopulated bin at the lowest He abundance value provided by the models (i.e., $Y = 0.23$), which occurs because this bin contains all the instances with $Y \leq 0.23$, instead of only the ones with $Y = 0.23$. This, in turn, also affects the age distribution, which can show multiple affected (overpopulated or underpopulated) bins. A regular normal fit to these distributions could thus be negatively impacted by these.

As a way to obtain reliable Gaussian fits when in presence of these spuriously over/underpopulated bins at the extremes of otherwise normal distributions, we propose to fit the corresponding cumulative distribution functions (CDFs) with Gaussian CDFs, instead. 
Similarly to fitting the regular distributions, these CDF fits have the mean and standard deviation of the $Y$ and age distributions as their fitting parameters. These CDFs are shown in the lower panels of Fig.~\ref{fig:iterations_rel}, with the best-fitting Gaussian CDF depicted with a dashed blue line. Additionally, using the parameters of the best-fitting CDF, we craft a Gaussian and compare it to the distributions in the upper panels of this figure. 
In the $Y$ distributions, we expect these CDF fits to be more robust in view of the fact that they are unaffected by the overpopulated bin. 
On the other hand, the age CDF can show over/underpopulated (older) bins, but it will always have one less bin affected than the regular distribution. Furthermore, even though we may not consider one or more age bins for the CDF fit, the instances within these are still considered in the bins with lower ages (and thus, in the fit). With these factors in mind, we also expect the CDF fit to the ages to be more robust.

\begin{figure}
    \centering
    \includegraphics[width=1\columnwidth]{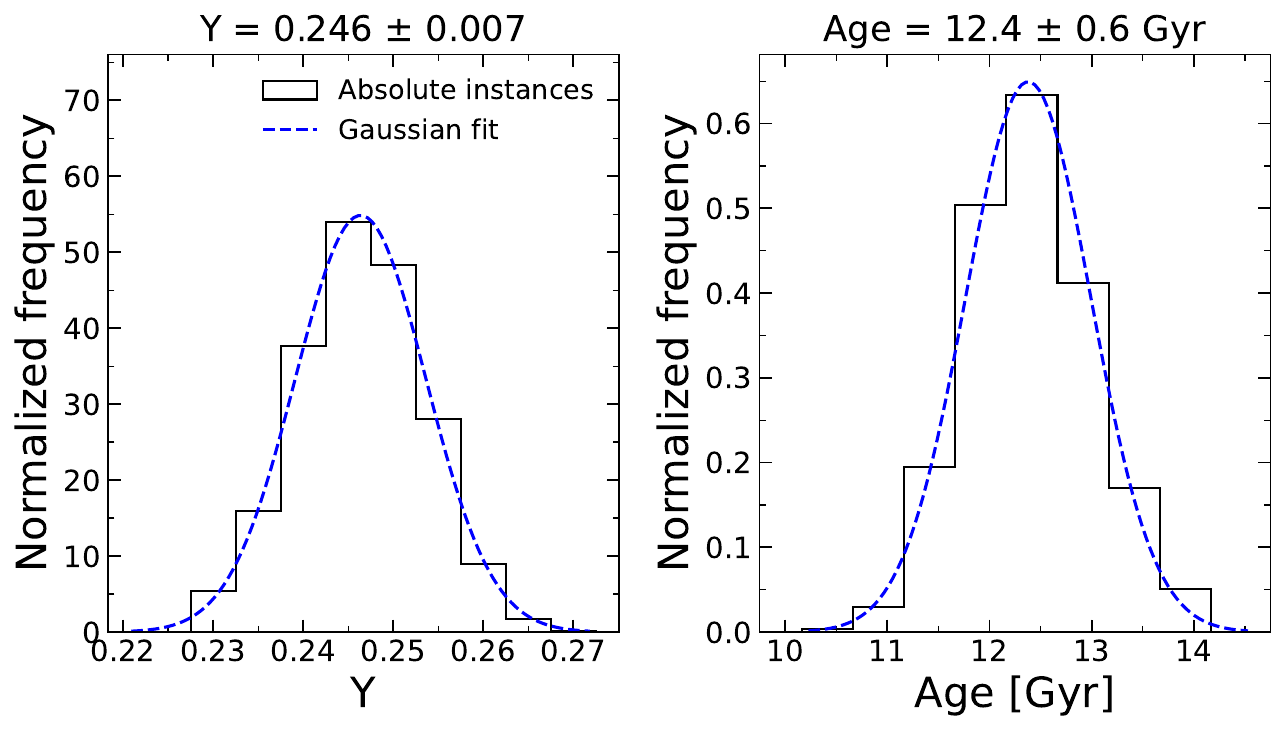}
    \caption{$Y$ (left) and age (right) distributions across $2000$ MC instances, using absolute PGPUC tracks with $Z = 0.006$. The best-fitting Gaussian distributions are shown as dashed blue lines, with their mean and standard deviation displayed in the title of the panels.}
    \label{fig:iterations_abs}
\end{figure}

The width of the bins in each histogram is selected following the Freedman-Diaconis rule \citep{Freedman1981}, with a minimum bin width in $Y$ and age of $0.005$ and $0.5$~Gyr, respectively, based on the resolution of our model grids. 
Figure~\ref{fig:iterations_abs} shows the $Y$ and age distributions obtained when using the absolute PGPUC models with $Z = 0.006$. It is clear that the absolute approach obtains much tighter distributions. In this case, it is also not necessary to employ the CDF to obtain reliable normal fits to the data. 
In both approaches, we adopt the nominal value and error in $Y$ and the age of the binary as the mean and standard deviation of the corresponding Gaussian fit.

\subsection{Precision of the relative approach}
\label{sec:precision}

As shown in Figs.~\ref{fig:iterations_rel} and \ref{fig:iterations_abs}, the statistical errors in $Y$ and age obtained with the relative approach are much larger than those obtained with the absolute approach, which holds true for both SECs. This arises from the fact that the absolute approach simultaneously fits two tracks to two stars, while the relative approach only fits one (relative) track to one (relative) star, thus reducing the fit's ability to precisely pinpoint the values of the parameters ($Y$, age). Additionally, Figs.~\ref{fig:absolute_tracks} and~\ref{fig:relative_tracks} show that the relative approach may be intrinsically less sensitive to variations in $Y$ than the absolute approach, which contributes to the larger (statistical) error bars obtained by the former.

Note, however, that our calculations use relative errors that arise from propagating the errors of the absolute parameters in quadrature. Such errors typically involve both a statistical and a systematic component. As already mentioned in Sect.~\ref{sec:introduction}, and indeed extensively exploited in the study of solar twins, the use of relative parameters may lead to a reduction in the impact of systematic uncertainties, as those should affect both stars similarly. 
It is of considerable interest, therefore, to analyze the impact of a reduction in the error of the differential parameters of the system, as compared with the propagated nominal errors that we have been using up to this point in our analysis. 

\begin{figure}
    \centering
	\includegraphics[width=0.9\columnwidth]{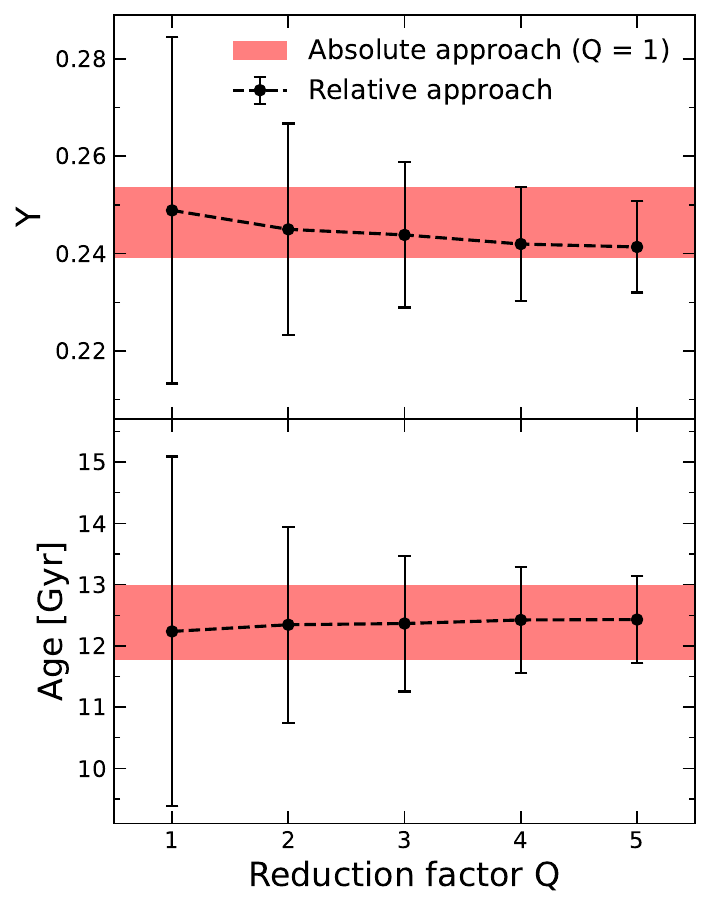}
    \caption{Random errors in $Y$ and age obtained with the relative approach (black error bars), as the errors in $L_{\text{bol}}$, $M$, $R$, and $g$ are reduced by a factor $Q$. The uncertainties obtained with the absolute approach (with unchanged errors) are displayed as red zones. All instances were computed using PGPUC models with $Z = 0.006$ and [$\alpha\text{/Fe]} = +0.3$.}
    \label{fig:error_reduction}
\end{figure}

To accomplish this, we arbitrarily shrink the errors in the relative $L_{\text{bol}}$, $M$, $R$, and $g$ values of the binary by a reduction factor $Q$, and we follow the procedure detailed in the previous sections to estimate anew the age and $Y$ of the binary, along with their respective errors. 
The results of this experiment are shown in Fig.~\ref{fig:error_reduction}, where we compare the precision obtained with the relative approach, taking a range of $Q$ values, to the precision of the absolute approach (with unchanged errors in the absolute parameters). As expected, the relative approach obtains much larger error bars than the absolute analysis when using the propagated errors; however, error bars of the same order as in the absolute approach can be inferred if the uncertainties in the relative $M$, $L_{\text{bol}}$, $R$, and $g$ of the binary are smaller than their propagated nominal values.

\subsection{Testing our methodology}
\label{sec:testing}

We have tested our methodology also by means of synthetic TO stars, generated using PGPUC isochrones, with physical parameters and associated error bars resembling those of V69's components (Table~\ref{tab:V69}). We explore synthetic stars with $Y$ and age values in the ranges of $0.24-0.32$ and $8-13$~Gyr, respectively. Then, we run the algorithm described in the previous sections to try and recover the $Y$ value and age of the original isochrones. 
Additionally, to further explore the robustness of both approaches, we repeat these tests with manipulated $T_{\text{eff}}$ values for both synthetic stars. We run tests with $T_{\text{eff}}$ values increased by $100$~K, and tests with those decreased by $100$~K. Then, we recalculate their radii and surface gravity, with these manipulated temperatures, and attempt to retrieve the $Y$ value and age of the isochrones.
This simulates the effect of the systematic uncertainty present in color-temperature transformations, which (among other factors) commonly affects the comparison between stellar models and empirical data. We choose to change the temperatures of both stars equally because these are very similar stars (see Table~\ref{tab:V69}), which should thus be affected in a similar manner by such systematics.

We perform these tests in both the absolute and relative approaches. In the latter case, we use two settings, one with propagated absolute errors ($Q = 1$), and another with reduced errors ($Q = 3$). This is done in order to ascertain whether more precise estimates of the relative parameters can lead to a more accurate $Y$ and age determination (in addition to a more precise one, as shown in Sect.~\ref{sec:precision}).

\begin{figure*}
    \centering
	\includegraphics[width=2\columnwidth]{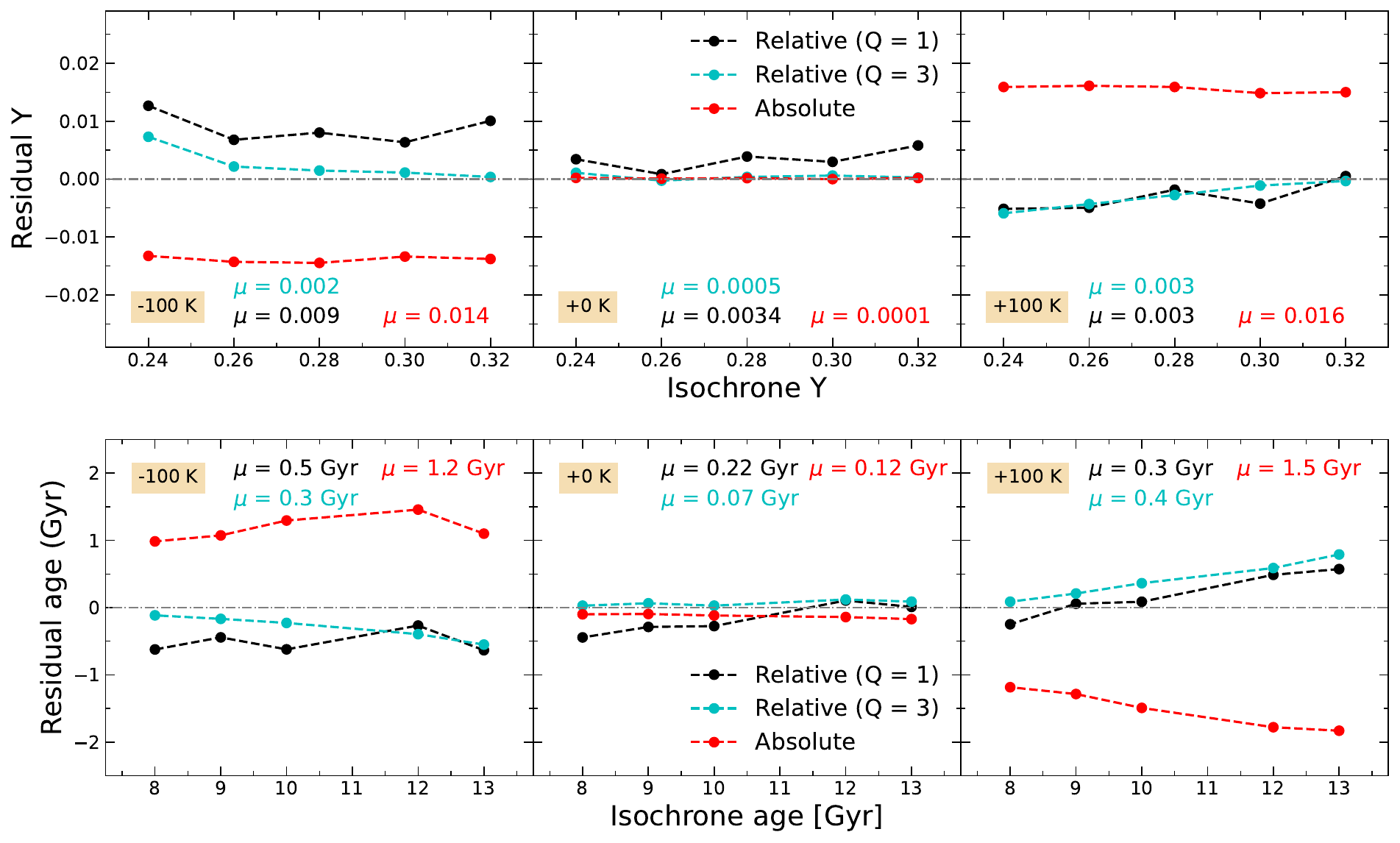}
    \caption{Residuals obtained in $Y$ (upper panels) and age (lower panels) when testing our methodology with synthetic TO stars (see Sect.~\ref{sec:testing}). Tests with decreased temperatures are shown in the left panels, while those with unchanged and increased temperatures are shown in the middle and right panels, respectively. 
    Red symbols represent the absolute approach, whereas the relative approach is indicated with black (propagated nominal errors) and cyan (errors reduced by a factor $Q=3$) symbols, respectively.
    Additionally, the mean values of the residuals are shown in their respective panel, with the corresponding color for each approach. These mean values are obtained using the absolute values of the residuals. The errors in each $Y$ and age estimation are not shown here, as these are discussed in Sect.~\ref{sec:precision}.}
    \label{fig:testing}
\end{figure*}

The results of these tests are depicted in Fig.~\ref{fig:testing}. Here, we also show the mean residuals (which are calculated using the absolute values of the residuals) each approach obtains in each of the three settings (i.e., decreased, increased, or unchanged temperatures). 
As shown here, when using unchanged temperatures, both the absolute and the relative method provide accurate age and $Y$ measurements, with the relative method showing better accuracy when using reduced errors.
When using manipulated temperatures (mimicking systematic errors), however, the absolute case shows large offsets in the age and $Y$ estimates. In this case, the He abundance is systematically under/over estimated (by about $0.015$, on average) when the temperatures are decrease/increased, and vice-versa for the ages (by about $1.4$~Gyr, on average). On the other hand, the relative approach shows much smaller offsets in the estimated age and $Y$ values than does its absolute counterpart, especially so when using reduced errors. These results support our hypothesis that the impact of systematic errors affecting the stellar parameters are reduced, when dealing with the binary components differentially.

\subsection{Correlation between the He abundance and age}
\label{sec: correlation}

\begin{figure*}
    \centering
	\includegraphics[width=2\columnwidth]{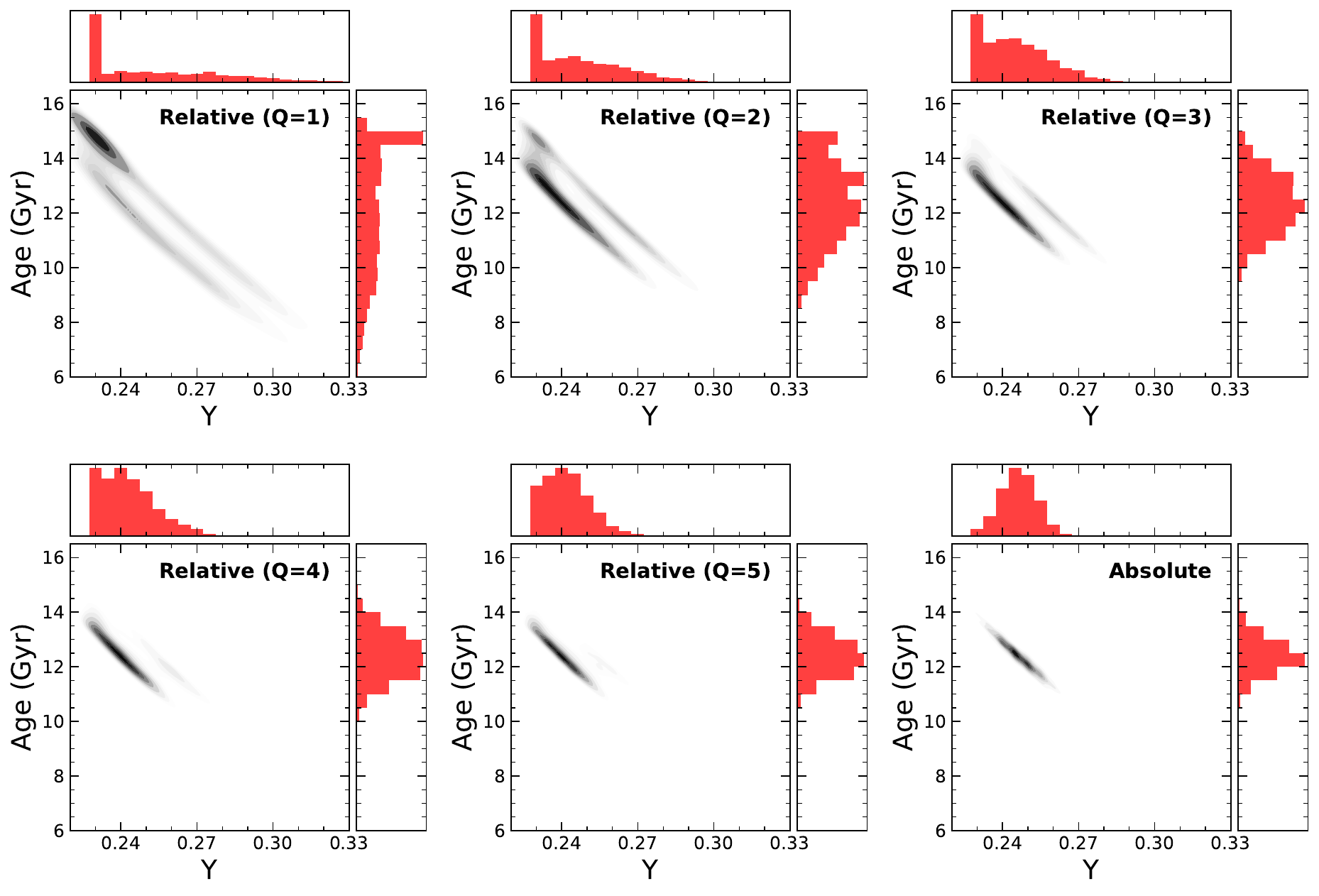}
    \caption{Contour plot of the $Y-{\rm age}$ distributions obtained in the MC instances. In the relative case, the distributions were obtained assuming increasing $Q$ values, from $Q=1$ (upper left) to $Q=5$ (bottom middle). The absolute case is depicted in the bottom right panel. In all cases, PGPUC models with $Z = 0.006$ were used. $Y$ and age histograms are included in each panel.}
    \label{fig:correlations}
\end{figure*}

As expected from theoretical models \citep[see, e.g.,][]{Thompson}, the inferred age of a binary system strongly depends on the adopted He abundance, with larger $Y$ leading to younger ages (due to faster evolution). Figure~\ref{fig:correlations}, where the $Y-{\rm age}$ distributions implied by our MC runs are shown, confirms this, both in the absolute and relative approaches.

Interestingly, Fig.~\ref{fig:correlations} also reveals a bimodality in the $Y-{\rm age}$ distribution when studying the system's relative parameters. 
This occurs because, in the relative case, the masses of the (absolute) tracks used to compute the best-fitting relative models (of the MC instances) tend to be concentrated near the $3 \sigma$ edges of the absolute mass distributions. Therefore, two peaks are formed in the distributions of these masses, which appear as the two ``ridges'' in Fig.~\ref{fig:correlations}.
This implies that, for systems with properties similar to V69's, it may not be straightforward to infer accurate masses from the peak of the relative MC solutions, as the (globally) most appropriate solutions~--- i.e., those with ${\rm MTSD} < 1$~--- are not concentrated around any specific mass values within the $3\sigma$ ranges.
In this sense, we note that the fraction of relatively poor solutions~--- i.e., those with ${\rm MTSD} > 1$~--- is higher in both ridges than it is in the ``valley'' region in between. Specifically, in the lower and upper ridges, 55\% and 24\% of the solutions are characterized by ${\rm MTSD} > 1$, whereas in the valley, this fraction is reduced to 20\%.

\section{Deriving the He abundance and age of V69}
\label{sec:results}

As discussed in Sect.~\ref{sec:testing}, the relative approach gives more accurate results when the uncertainties are reduced. 
Therefore, in what follows we adopt errors reduced by a factor $Q = 3$ to estimate V69's $Y$ value and age. Whether such a reduction in the errors of one or more relative parameters can be achieved empirically is beyond the scope of this paper, though we note that some studies have shown that at least some reduction is indeed possible \citep{Southworth2007, Torres2014, Taormina2024}.

\begin{figure}
    \centering
	\includegraphics[width=0.9\columnwidth]{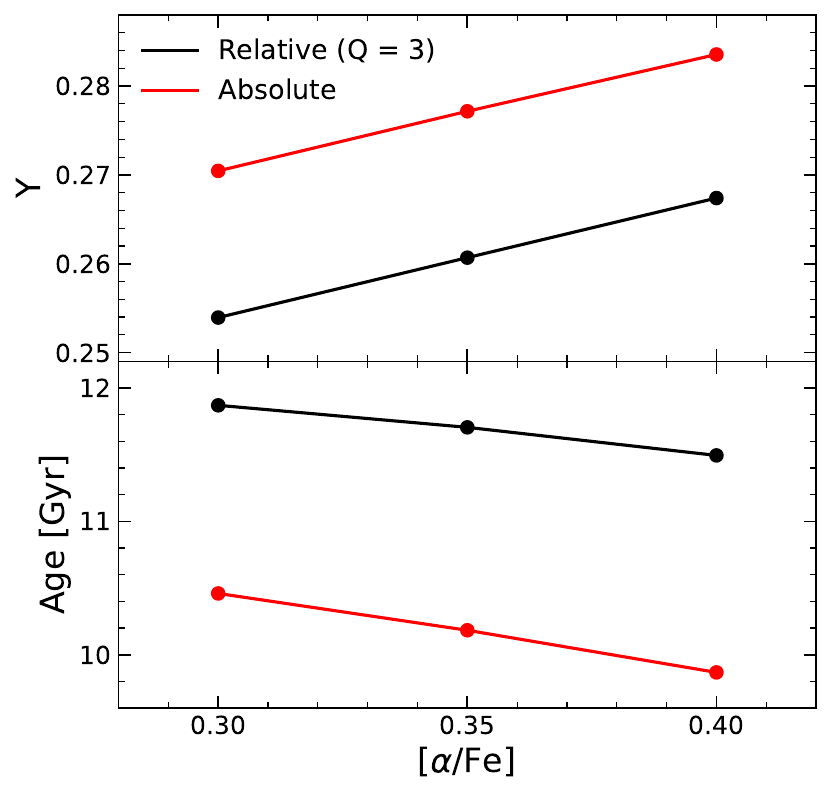}
    \caption{He abundance (upper panel) and age (lower panel) as a function of [$\alpha\text{/Fe]}$ for the relative (black lines) and absolute (red lines) approaches. These calculations make use of VR models with $\text{[Fe/H]} = -0.6$, and [$\alpha\text{/Fe]}$ in a range of $0.3 - 0.4$. The errors in each $Y$ and age estimation are not included; these are discussed in Sect.~\ref{sec:precision}.}
    \label{fig:aFe_dependency}
\end{figure}

Due to the limitations of the grid provided by \citet{PGPUC2012}, we use PGPUC models with [$\alpha\text{/Fe]} = +0.3$. However, the adopted $\alpha$-element enhancement of 47 Tuc is $+0.4$. Thus, in order to obtain age and $Y$ estimates with the adopted chemical composition of 47 Tuc ([$\alpha\text{/Fe]} = +0.4$ and $\text{[Fe/H]} = -0.71$), we determine the effect of differing [$\alpha\text{/Fe]}$ values on our age and He abundance estimations. 
This is shown in Fig.~\ref{fig:aFe_dependency}, where we present our results with both approaches, using VR models with [$\alpha\text{/Fe]}$ values in the range of $0.3 - 0.4$, and a fixed $\text{[Fe/H]} = -0.6$.
We find that, in either approach, an increase in the assumed $\alpha$-element enhancement of the binary system correlates with a decrease in its estimated age and an increase in its He abundance.
Figure~\ref{fig:aFe_dependency} shows that, in the relative approach, an increment in [$\alpha\text{/Fe]}$ of $0.1$ leads to an increase in $Y$ of roughly $0.013$ and a decline in age close to $0.4$~Gyr. On the other hand, in the absolute approach, an equal increment in [$\alpha\text{/Fe]}$ enlarges the estimated $Y$ by $0.013$, and reduces the age by $0.6$~Gyr.
Thus, both approaches produce age and $Y$ estimations with very similar correlations with [$\alpha\text{/Fe]}$, with the relative analysis providing a slightly weaker age dependency.

\begin{figure}
    \centering
	\includegraphics[width=0.9\columnwidth]{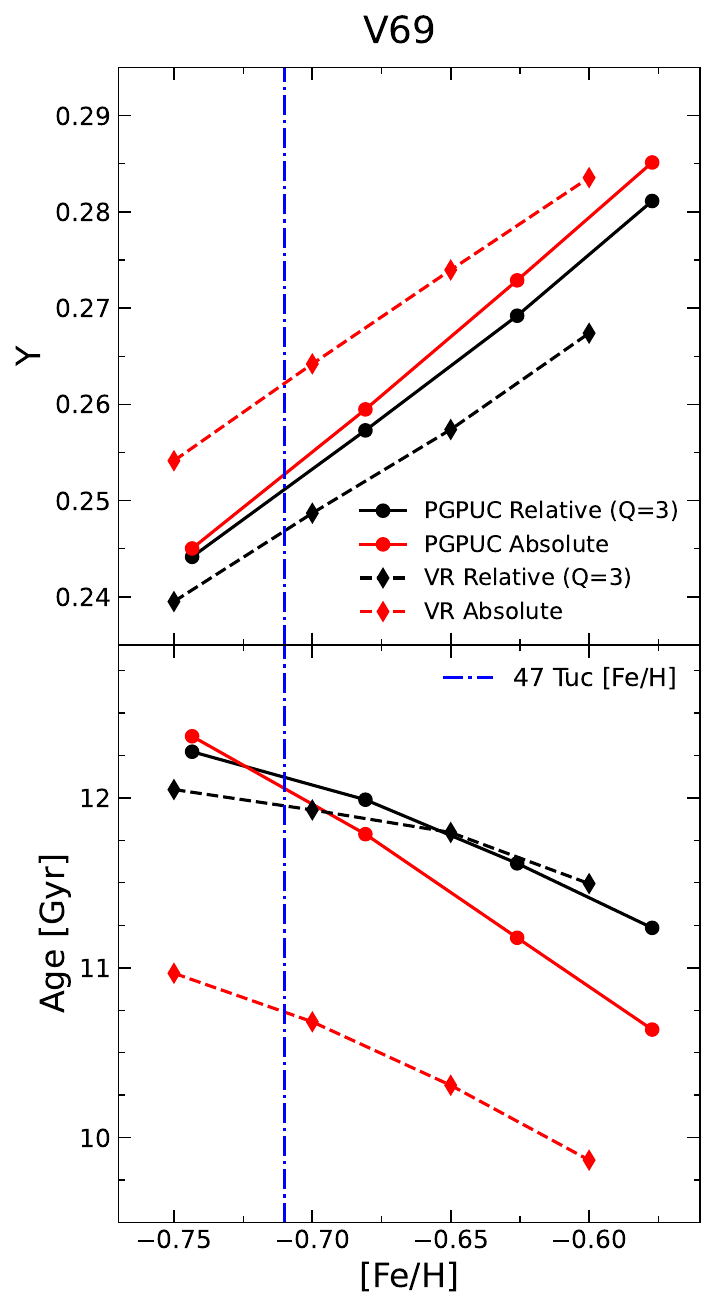}
    \caption{He abundance (upper panel) and age  (lower panel) as a function of [Fe/H] for the relative (black lines) and absolute (red lines) approaches, using PGPUC (solid lines) and VR (dashed lines) models. These results use [$\alpha\text{/Fe]} = +0.4$. The assumed [Fe/H] value of V69 is depicted as a blue dash-dotted line. The errors in each $Y$ and age estimation are not included, as these are discussed in Sect.~\ref{sec:precision}.}
    \label{fig:Z_dependency}
\end{figure}

We apply these slopes as corrections to the PGPUC results obtained with [$\alpha\text{/Fe]} = +0.3$, in order to shift these to estimations with [$\alpha\text{/Fe]} = +0.4$. Figure~\ref{fig:Z_dependency} depicts the sensitivity of the measured ages and He abundances to the [Fe/H] of the models, using both PGPUC and VR models. All results shown correspond to those obtained with [$\alpha\text{/Fe]} = +0.4$. 
Note that Fig.~\ref{fig:Z_dependency} does not include the errors in each $Y$ and age estimation; however, these are of the order of those shown in Fig.~\ref{fig:error_reduction} (with $Q = 3$, in the relative approach), and do not change in any significant way with [Fe/H].

We find that an increase in the assumed metal content of the binary system correlates with a decrease in its estimated age and an increase in its He abundance, akin to the correlations our results show with [$\alpha\text{/Fe]}$. These trends are consistent across both approaches. However, the relative results show a weaker dependence with [Fe/H] than the absolute measurements, particularly in age.
Additionally, the PGPUC results for both $Y$ and age show a very well-defined linear dependency on [Fe/H], while the relative VR results display a mildly disjointed age dependency at lower [Fe/H]. This occurs because we can find a $Y$ value outside (or very close to the limit of) the range of He abundances provided by the models (in the VR case, $Y = [0.25, 0.33]$), which can lead to a slight overestimation and underestimation of the $Y$ value and age, respectively. This happens for the relative calculations performed using the VR models with both $\text{[Fe/H]} = -0.75$ and $-0.7$.

As shown in Fig.~\ref{fig:Z_dependency}, 
PGPUC and VR tracks lead to inferred $Y$ values that are within about $0.01$ of each other, with the relative approach implying slightly lower He abundances. 
On the other hand, it is clear that the relative approach provides age estimations that are less model-dependent than the absolute approach. The former produces ages that agree to within about $0.2$~Gyr when comparing PGPUC and VR results (even when some age estimates obtained using VR models are slightly underestimated, as discussed in the previous paragraph), while the latter provides age estimates which differ by at least $1$~Gyr between PGPUC and VR.

Explaining this result is outside the scope of this paper, but it is again evidence that systematic effects can have a deleterious impact on the determination of GC $Y$ and age from binary TO stars. 
Differences in the input physics adopted in the PGPUC and VR codes could be one contributing factor. For instance, PGPUC adopts the FreeEOS equation of state (EOS) \citep{Irwin2012}, whereas VR uses a somewhat more rudimentary EOS \citep{VandenBerg2012,VandenBerg2014}. At V69's metallicity, this may be an important contributor to the $\approx 65$~K offset that we find between PGPUC and VR tracks at the TO level, the latter set being systematically cooler than the former (Fig.~\ref{fig:absolute_tracks}). Though we do not use temperatures directly in our method, we do use them to obtain gravities and radii, both of which are parameters used in our goodness-of-fit diagnostics (see Eq.~\ref{eq:TSD}). Other relevant ingredients that could affect the results of different SECs, and the placement of their evolutionary tracks and isochrones in color-magnitude diagrams (CMDs), include the adopted outer boundary conditions, mixing-length formalism adopted for the treatment of convection, low-temperature radiative opacities, color-temperature relations and bolometric corrections, and reference (solar) abundance mix, among others \citep[see, e.g.,][and references therein]{Gallart2005,Catelan2007,Catelan2013,Pisa2013b,Cassisi2016,VandenBerg2016}.

Such a temperature offset in the models is akin to the tests we describe in Sect.~\ref{sec:testing}. Particularly, using these VR models, which are systematically cooler, should be roughly equivalent to the tests where we increase the temperature of the stars (shown in the right panels of Fig.~\ref{fig:testing}), albeit at a reduced level (we adopted a temperature offset of $100$~K in those tests, whereas the difference in temperatures between PGPUC and VR tracks at the TO level is about $65$~K instead). 
Nevertheless, the similarities between the results shown in Fig.~\ref{fig:Z_dependency} and the tests using increased temperatures are evident. In the absolute case, the VR models (as compared to PGPUC ones) and these tests alike lead to significantly underestimated ages and overestimated He abundances. In the relative approach, the differences are much reduced, but still fully consistent with the trends seen in Fig.~\ref{fig:Z_dependency}, particularly considering the slightly lower ages obtained with VR models. 

With the relative approach, we find that the age and He abundance of V69, and thus of 47~Tuc, are $12.1 \pm 1.1$~Gyr and $0.247 \pm 0.015$, respectively. These uncertainties correspond to the statistical errors discussed in Sect.~\ref{sec:precision}, with $Q = 3$. 
The {\em real} statistical uncertainties depend on how much more accurate a direct determination of the relative parameters is, compared to the propagated errors in the individual components' separately measured parameters. In addition, the possibility of correlated uncertainties in these parameters, which was not considered in this work, should be evaluated and taken into account if necessary as well.
In any case, it should be noted that the relative method becomes less sensitive, the more similar the two binary components are; indeed, in the limit where both stars are identical, the method cannot be applied. In the case of V69, as we have seen earlier, both stars are indeed quite similar, and are both currently located at 47 Tuc's TO point. In this sense, binary pairs in which the stars differ more in their properties, with one of the components being somewhat less evolved than the other, would help beat down the statistical uncertainties, in the relative approach. This, in fact, is the case of E32 in 47~Tuc, a DEB system which we plan to address in a forthcoming paper. 

These age estimates are in good agreement with the results presented in \citet{Brogaard2017} and \citet{Thompson2020}, who find that the age of V69 is $11.8 \pm 1.5\, (3\sigma)$~Gyr and $12.0 \pm 0.5\, (1\sigma)$~Gyr, respectively. These were obtained by comparing the absolute parameters of V69 with stellar models, combined with isochrone fitting to the CMD of 47 Tuc. Additionally, to obtain the age of 47 Tuc, \citet{Thompson2020} analysed not only V69, but also E32.

At first glance, these tight constraints on the age of V69 could indicate that our method, which obtains much looser constraints, is not an effective age indicator. However, both quoted papers rely on a single SEC; thus, their quoted errors do not consider the systematic uncertainty that arises from using different sets of evolutionary tracks. This systematic, as discussed in Sect.~\ref{sec:results}, can be most significant when working with the absolute parameters of the stars. This is also shown in Table~9 of \citet{Thompson}, who estimated V69's age using five different SECs and different combinations of physical parameters, including mass, radius, and luminosity. 
They find ages that differ by up to $2$~Gyr, depending on the models used and the adopted He abundance (as they do not simultaneously derive the latter along with the age). 

Furthermore, \citet{Brogaard2017} and \citet{Thompson2020} find an approximate $Y$ value for V69 by performing isochrone fitting to the CMD of the cluster, which they then use to estimate the age of the binary. As both works point out, this assumes that V69 (and E32, in the \citeauthor{Thompson2020} case) belongs to a specific population of 47 Tuc.
In contrast, our methodology neither assumes a $Y$ value a priori, nor does it require isochrone-fitting to be performed to the cluster's CMD. 
Furthermore, the method proposed in this work could be used to test these assumptions and study the He abundance of the multiple populations in GCs, in the event different binaries are found belonging to different subpopulations with different $Y$ values within a given cluster. Considering these factors, we are confident that the constraints we have placed on the $Y$ value and age of 47 Tuc are robust, and that the method can be competitive in the search for He abundance and age variations, including within individual star clusters.

\section{Final remarks}
\label{sec:conclusions}

In this paper, we have proposed and tested a new approach to simultaneously measure the age and He abundances of star clusters. It relies on the differential evolutionary properties of DEBs in which at least one of the components is close to the main-sequence TO. We argue that the method is less sensitive to systematic effects than methods relying on the absolute physical parameters of the stars, and demonstrate that it can give useful results in the case of the DEB V69 in 47 Tuc. As a proof of concept, we analyze the system assuming that the errors in the relative parameters can be reduced by a factor of 3 compared to the propagated formal errors in the individual, absolute parameters of the components. In this way, we are able to infer an age for the system of $12.1 \pm 1.1$ Gyr and a He abundance of $0.247 \pm 0.015$. Naturally, the actual errors in $Y$ and age that can be achieved with our method depend critically on how precisely each of the several relative parameters of the binary can be determined empirically. Application of the method to other suitable DEB systems, as well as further exploration of the methodology, are strongly encouraged. 
In particular, efforts at determining the {\em relative} parameters of DEB stars, with a decreased sensitivity to systematic effects that would otherwise affect each component's {\em absolute} parameters individually, would prove especially useful in putting the proposed methodology on a firmer footing.

\begin{acknowledgements}
We gratefully acknowledge the constructive comments and suggestions provided by the referee. Support for this project is provided by ANID's FONDECYT Regular grants \#1171273 and 1231637; ANID's Millennium Science Initiative through grants ICN12\textunderscore 009 and AIM23-0001, awarded to the Millennium Institute of Astrophysics (MAS); and ANID's Basal project FB210003. NCC acknowledges support from SOCHIAS grant ``Beca Adelina''. We are very grateful to Don VandenBerg for providing us with his evolutionary tracks, and programs to interpolate within these. 

The research for this work made use of the following Python packages: Astropy \citep{Astropy2013, Astropy2018}, Matplotlib \citep{Matplotlib}, NumPy \citep{Numpy}, SciPy \citep{Scipy}, Jupyter \citep{Jupyter}, and pandas \citep{Pandas}.
\end{acknowledgements}

\bibliographystyle{aa}
\bibliography{bibliography}

\begin{appendix}

\section{Track interpolation}
\label{app:interpolation}

In order to interpolate evolutionary tracks from the VR isochrone grids, we compute a linear interpolation of their parameters ($T_{\text{eff}}$, $L_{\text{bol}}$, $R$, $g$) as a function of the mass values associated with each point along the isochrone, and we evaluate these interpolations on the mass of the original track we seek to reproduce. 
The resulting track has $135$ evolutionary points (the number of isochrones used in the interpolation), with minimum and maximum ages of $1$~Gyr and $18$~Gyr, respectively. To improve the time resolution of the track, we perform a second interpolation of the resulting track parameters as a function of the evolutionary age, using the INTEP interpolation subroutine \citep{Hill1982}.

\begin{figure*}
    \centering
	\includegraphics[width=1.75\columnwidth]{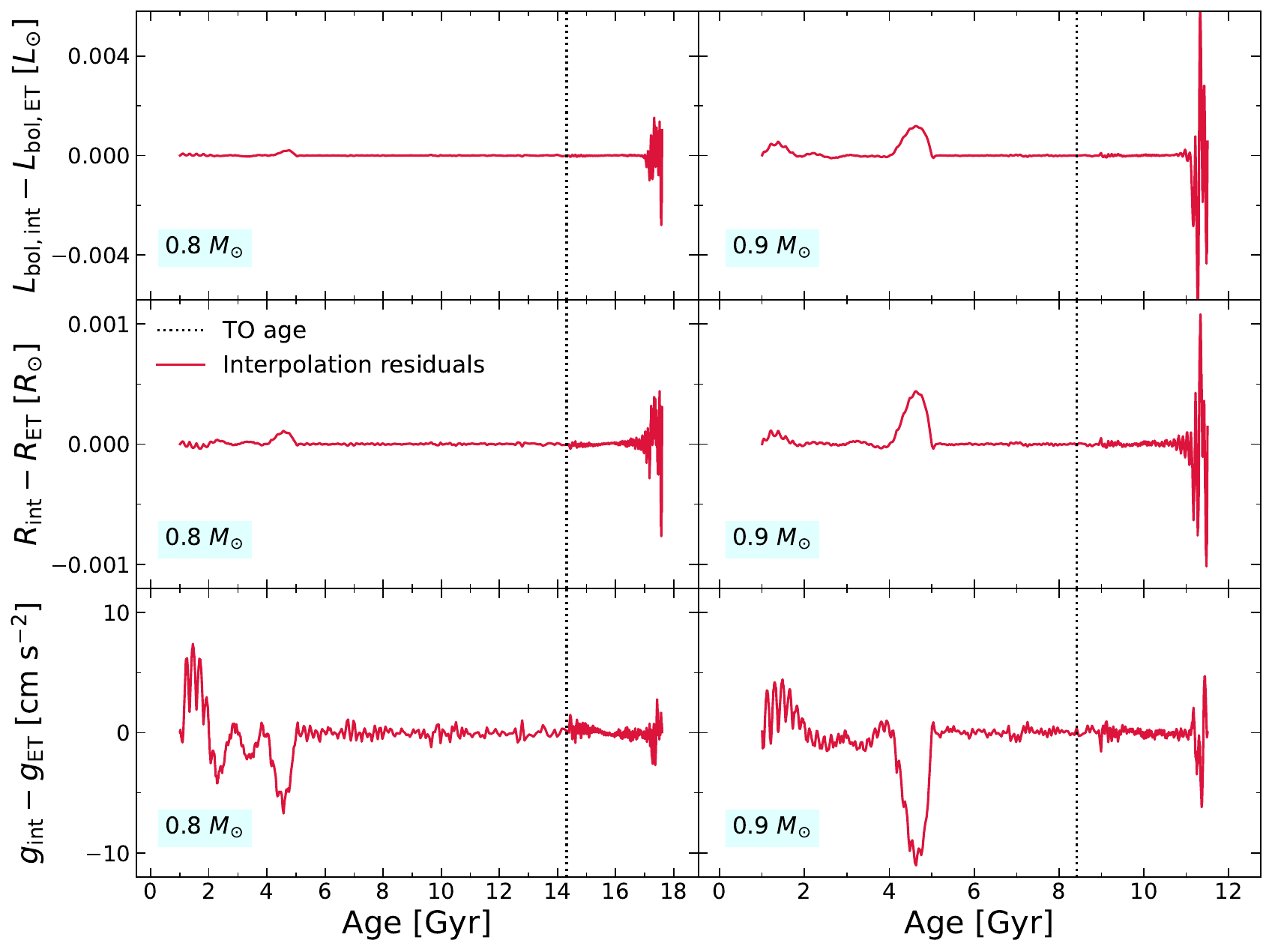}
    \caption{Residuals of the VR interpolations in $L_{\text{bol}}$ (upper panels), $R$ (middle panels), and $g$ (lower panels) as a function of the evolutionary age. The tracks were computed with a chemical compositon described by $Y = 0.25$, $[\alpha/\text{Fe}] = 0.3$, and $[\text{Fe/H}] = 0.7$, and correspond to masses of $0.8 \, M_{\odot}$ (left panels) and $0.9 \, M_{\odot}$ (right panels). The evolutionary ages shown cover the evolution of the star up until the base of the RGB. The TO age of the tracks is shown in each panel as a vertical dotted line.}
    \label{fig:app-residuals}
\end{figure*}

To judge the accuracy of this interpolation method,
we compare two tracks (with $M = 0.8$ and $0.9 \, M_{\odot}$), interpolated from our VR isochrone grid (Table~\ref{tab:Tracks}), to tracks with the same masses (and chemical compositions), provided directly by the VR interpolation software \citep{VandenBerg2014}. The chemical composition of the tracks is $[\text{Fe/H}] = 0.7$, $Y = 0.25$, and $[\alpha/\text{Fe}] = 0.3$. To compare these tracks, we plot their residuals in $L_{\text{bol}}$, $R$, and $g$, as a function of the evolutionary age along the track. The results of this test are shown in Fig.~\ref{fig:app-residuals}. The evolution of the tracks is shown up until they enter the RGB stage, since our main focus is the TO point of the tracks, not their later evolution. 
This test shows that the interpolation reproduces the expected tracks with very good precision in the MS and TO stages, showing minor bumps in the residuals before $5$~Gyr, which occur due to the larger time-step of our grid (Table~\ref{tab:Tracks}) in these early ages. 
The interpolation starts to break down in the late stages of the subgiant branch. This occurs because our pre-INTEP tracks, with $135$ evolutionary ages, have thinly populated post-MS stages, leading to problems with the interpolation. Nevertheless, this does not represent a problem for our analysis, given that we are dealing with two stars located very near the TO point. 

\section{On the use of parameter ratios}
\label{app:ratios}

As mentioned in Sect.~\ref{sec:diff_params}, parameter {\em ratios} may constitute an interesting alternative to the use of differential parameters, motivated by the fact that some such ratios can be measured more straightforwardly \citep[e.g.,][]{Popper1985, Maxted2016, Morales2022}.

\begin{figure*}
    \centering
	\includegraphics[width=2\columnwidth]{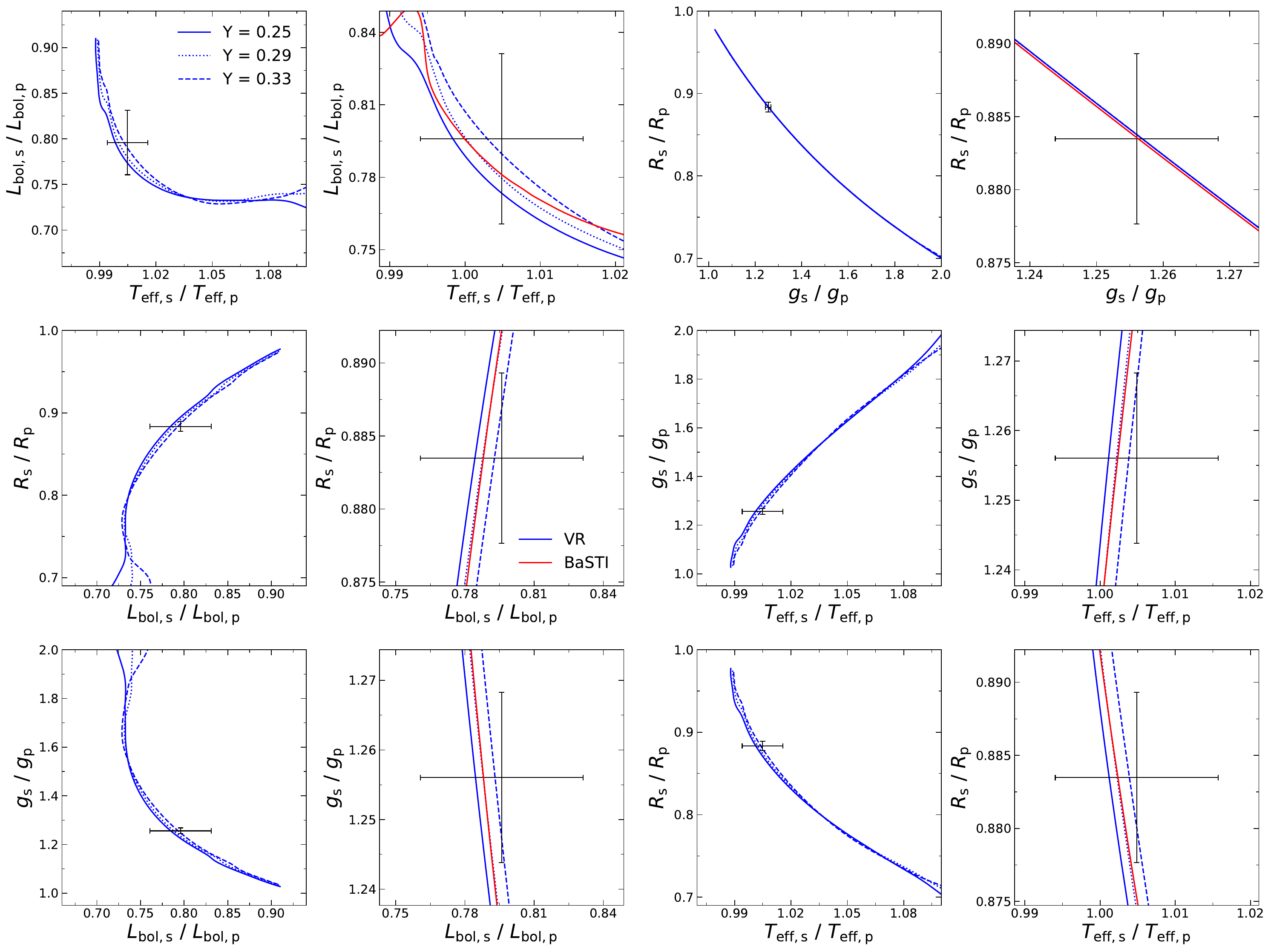}
    \caption{Similar to Fig.~\ref{fig:absolute_tracks}, but displaying ratios of parameters, instead of the absolute ones, for different combinations of $L_{\rm bol,s}/L_{\rm bol,p}$, $R_{\rm s}/R_{\rm p}$, $g_{\rm s}/g_{\rm p}$, and $T_{\rm eff,s}/T_{\rm eff,p}$. 
    The nominal masses and other empirical parameters of V69's components are the same as listed in Table~\ref{tab:V69}; since the absolute mass values are fixed, tracks for different $Y$ values have different TO ages, as also happens in the case of Fig.~\ref{fig:relative_tracks}.
    The first and third columns show the same evolutionary stages as those covered by the relative tracks presented in Fig.~\ref{fig:relative_tracks}. The second and fourth columns are zoom-in plots corresponding to the first and third columns, respectively. In each panel, blue solid, dotted, and dashed lines correspond to VR tracks for $Y = 0.25$, 0.29, and 0.33, respectively (Table~\ref{tab:Tracks}), whereas red tracks in the zoomed-in plots correspond to BaSTI tracks with $Y = 0.255$, $Z = 0.006$.     
    In this case, we only show VR tracks (for easier viewing), but the PGPUC models show the same behaviour. The point with error bars correspond to the empirical parameters derived for V69 from the empirical data (Table~\ref{tab:V69}), with the error bars being propagated from the absolute values.}
    \label{fig:ratios}
\end{figure*}

To explore this possibility, in Fig.~\ref{fig:ratios} we recreate Fig.~\ref{fig:relative_tracks}, but plotting parameter ratios instead of differences. In this figure, the empirical measurements (with propagated errors) for 47~Tuc's V69 DEB system are compared with VR evolutionary tracks for the indicated $Y$ values. PGPUC tracks, not shown for the sake of clarity, display essentially the same behavior, as also seen in Fig.~\ref{fig:relative_tracks}. In the zoomed-in panels of Fig.~\ref{fig:ratios}, we also add BaSTI tracks with [Fe/H] = $-0.7$, $Y = 0.255$, [$\alpha$/Fe] = $0.4$, and $Z = 0.006$, which show excellent agreement with the VR (and thus PGPUC) tracks for a similar chemical composition.

Figure~\ref{fig:ratios} reveals, intriguingly, that, irrespective of the theoretical models adopted, the parameter ratios consistently require models with what appear to be unrealistically high $Y$ values to match the empirical data for V69. We have checked that the same happens in the case of 47~Tuc's E32 DEB system. This suggests that, unlike differential parameters, parameter ratios are sensitive to systematic errors, and thus less amenable to a method such as proposed in this paper. 
Additionally, these ratios show a weak dependency with $Y$, as all the tracks included in Fig.~\ref{fig:ratios} pass within the (propagated) error bars.

Further exploration of parameter ratios is strongly encouraged, in order to explain the intriguing discrepancy shown in Fig.~\ref{fig:ratios} and further explore their utilization in studies of the He abundance and age of DEB systems.

\end{appendix}

\end{document}